\documentstyle[aps,multicol,pre,epsfig]{revtex}

\newcommand{\be}{\begin{equation}}
\newcommand{\ee}{\end{equation}}
\newcommand{\lb}{\ell_B}

\newcommand{\Zeff}{Z_{\hbox{\scriptsize eff}}}
\newcommand{\Zbare}{Z_{\hbox{\scriptsize bare}}}
\newcommand{\Zsat}{Z_{\hbox{\scriptsize sat}}}

\newcommand{\lasat}{\lambda_{\hbox{\scriptsize sat}}}

\begin{document}
\title{Analytical estimate of effective charges at saturation in 
Poisson-Boltzmann cell models}
\author{Emmanuel Trizac$^{1}$, Miguel Aubouy$^{2}$ and 
Lyd\'eric Bocquet$^{3}$}
\address{$^{1}$ Laboratoire de Physique Th\'eorique, 
UMR CNRS 8627, B{\^a}timent 210, Universit{\'e}  Paris-Sud,
91405 Orsay Cedex, France, \\
$^{2}$ S.I.3M., D.R.F.M.C., CEA-DSM Grenoble, 
17 rue des Martyrs, 38054 Grenoble Cedex 9, France\\
$^{3}$ D\'epartement de Physique des Mat\'eriaux, UMR CNRS 5586, 
Universit\'e Lyon-I, 43 Bd du 11 Novembre 1918, 69622 Villeurbanne
Cedex, France 
}

\date{\today}
\maketitle

\begin{abstract}
We propose a simple approximation scheme to compute the effective charge
of highly charged colloids (spherical or cylindrical with infinite length). 
Within non-linear Poisson-Boltzmann theory, we start from an  
expression of the effective charge
in the infinite dilution limit which is asymptotically valid for large salt 
concentrations; this result is then extended to finite colloidal concentration,
approximating the salt partitioning effect which relates the salt content in 
the suspension to that of a dializing reservoir. 
This leads to an analytical expression of
the effective charge as a function of colloid volume fraction and salt concentration.
These results compare favorably with the effective charges {\em at saturation}
(i.e. in the limit of large bare charge) computed numerically following the
standard prescription proposed by Alexander {\it et al.} within the cell model. 
\end{abstract}



\section{Introduction}

In colloidal suspensions, the charge ratio between the poly-ions (colloids)
and the counter-ions or salt ions may be 
larger than 10$^4$. This considerable
asymmetry invalidates most of standard liquid state theories. The concept
of charge renormalization however allows to significantly simplify
the description of such systems: the electrostatic coupling between 
the poly-ions and oppositely charged micro-species induces a 
strong accumulation (or ``condensation'') near the poly-ion surface.
The idea is then to consider a colloid plus its ``captive'' micro-ions
as an entity carrying an effective (or renormalized) 
charge $\Zeff$, {\it a priori} much smaller 
than the bare charge $\Zbare$. Consequently, 
except in the immediate vicinity
of the colloids where linearization schemes fail, the interactions
in the suspension are well described by Debye-H\"uckel-like
linearized theories, provided that the bare charge is replaced by the 
renormalized one $\Zeff$. Schematically, the linear effects of screening
by an electrolyte induce a dressing of the bare Coulomb potential
[$\Zbare/r \to \Zbare \exp(-\kappa r)/r$ where $\kappa^{-1}$ is the Debye length,
see below], while the non-linear
effects of screening imply the identification 
$\Zbare \exp(-\kappa r)/r \to \Zeff \exp(-\kappa' r)/r$ (with possibly
$\kappa'\neq \kappa$ \cite{Kjellander}). 

Several reviews discussing the notion of charge 
renormalization have appeared
recently \cite{Kjellander,Belloni,Hansen,Levin}. In the colloid science
field this concept has been introduced by Alexander {\it et al.} 
\cite{Alexander} in the context of the Poisson-Boltzmann (PB) cell 
model, but had been widely accepted
since the fifties in the field of linear polyelectrolytes 
\cite{Katchalsky,Manning}. The definition of an effective charge
from the far field potential created by an isolated macro-ion in 
an electrolyte is unambiguous \cite{Belloni,Letter,Rque}, 
at least for the simple cases
of spherical or infinitely long cylindrical 
macro-ions we shall consider
here, even if the case of less 
symmetric poly-ions deserves more attention (see e.g. \cite{JPCM}). 
Within a cell model, introduced to replace the complicated
many body problem of colloids in solution 
by a simpler  one-particle system
\cite{Marcus,Trizac,Trizacbis,DesernoGrunberg}, the definition 
of an effective charge
is more elusive, but the celebrated proposal made by Alexander {\it et al.}
amounts to find the optimal linearized 
PB potential matching the non-linear one
at the cell boundary. 
The cell approach, which validity has been assessed by several studies
\cite{Hartl,Bitzer,Lowen,Gisler,Stevens,Lobaskin}, appears to provide
a reasonable description of solutions containing mono-valent counter-ions
\cite{Groot,Hribar,Evilevitch}. 
The effective charge is obtained by integrating the
charge density deduced from the linearized potential over the region
accessible to micro-ions, or equivalently from Gauss theorem at the colloid's 
surface. Note that a renormalization of $Z$ also implies a renormalization
of the screening constant $\kappa$. For low poly-ion bare charges $\Zbare$,
linearizing PB equation is a valid approximation 
so that $\Zeff \simeq \Zbare$,
whereas in the opposite limit of high bare charges, 
$\Zeff$ was found numerically
to saturate to a value independent of $\Zbare$:
$\Zeff\simeq \Zsat$. The saturation value $\Zsat$ depends on the
geometry of the colloid, the temperature or the quantity of added-salt.
Unfortunately, no analytical
prediction is available for these dependences.

In this letter, we propose an
approximate analytical expression for the effective charge of spherical or rod-like 
macro-ions, as a function of macro-ion density or salt content. 
Our approach starts with an estimate of the saturation value of the 
effective charge in the infinite dilution limit, as deduced from recent analytical
results \cite{Shkel}. In the case of finite colloidal dilution, the osmotic 
equilibrium of the suspension with a salt reservoir is modeled using
a Donnan equilibrium approximation.
This allows to derive a simple polynomial equation 
of degree 4 fulfilled by the
{\em saturation value} $\Zsat$ of the effective charge $\Zeff$,
in the case of a symmetric 1:1 electrolyte.
In the limiting case of no added salt, the above
equation is easily solved analytically and provides results
in quantitative agreement with the saturation effective charges
following Alexander's prescription.



\section{General framework and method}

We consider first the situation of an isolated macro-ion of given surface
charge density in a electrolyte of bulk density $n_0$ (no confinement).
The solvent is considered
as a medium of uniform dielectric (CGS) permittivity $\varepsilon$.
Within Poisson-Boltzmann theory, the electrostatic potential,
when assumed to vanish far from macro-ion 
obeys the equation
\be
\nabla^2 \phi = \kappa_0^2 \sinh \phi,
\label{eq:pb}
\ee
where the screening factor $\kappa_0$ is defined as
$\kappa_0^2 = 8 \pi \lb n_0$ and the 
Bjerrum length quantifies the strength
of electrostatic coupling: $\lb = e^2/(\varepsilon kT)$ ($e>0$ denotes
the elementary charge and $kT$ is the 
thermal energy). A complete asymptotic
solution of Eq. (\ref{eq:pb}) has been 
obtained recently by Shkel {\it et al.}
\cite{Shkel} in spherical and cylindrical geometry. 
From the far-field behaviour
of the corresponding solutions, one obtains after some algebra 
the effective charges at saturation
\begin{eqnarray}
&& \Zsat \frac{\lb}{a} = 4 \kappa_0 a + 6 + 
{\cal O}\left(\frac{1}{\kappa_0 a}\right) \quad \hbox{for spheres}
\label{eq:za}\\
&&\lasat \lb = 2 \kappa_0 a + \frac{3}{2} + 
{\cal O}\left(\frac{1}{\kappa_0 a}\right) \quad \hbox{for cylinders}.
\label{eq:zb}
\end{eqnarray}
In these equations, $a$ denotes the radius of the macro-ion under
consideration, $Ze$ the total charge in the case of spheres and
$\lambda e$ the line charge density in the case of cylinders (with
infinite length). Expressions (\ref{eq:za}) and (\ref{eq:zb}) are the
exact expansions of the saturation charges in the limit of large
$\kappa_0 a$ but are in practice accurate as soon 
as $\kappa_0 a >1$ (not shown).

Our goal is to translate these relations into expressions that would
approximate the effective charges in 
confined geometry, where the macro-ion
is enclosed in a cell \cite{Alexander}. 
To this aim, we first find an approximation for
the relevant screening factor 
$\kappa_\star$ before inserting
it into (\ref{eq:za}) and (\ref{eq:zb}) making the substitution
$\kappa_\star \leftrightarrow \kappa_0$. The implicit 
assumption is that the mean salt density in the cell 
is related to the effective charge in a similar manner 
as in the infinite dilution limit.

In confined geometry, PB equation
still takes the form (\ref{eq:pb}), where 
$\kappa_0=(8 \pi \lb n_0)^{1/2}$ should now be considered
as the inverse screening length in a (neutral) salt reservoir 
in osmotic equilibrium
with the solution, through a membrane permeable to micro-species ($n_0$ is 
thus now the salt density in the reservoir). 
It may be shown that the relevant screening factor 
in the cell is related to the micro-ions density at the cell boundary
\cite{Letter}:
\begin{equation}
\kappa^2_\star=4\pi \ell _{B}\left[\rho ^{+}(R_{WS})+\rho
^{-}(R_{WS})\right].  \label{Kdonnan}
\end{equation}
In this equation $R_{WS}$ is the radius of the Wigner-Seitz (WS) confining cell.
We have recently proposed an efficient prescription to compute
$\kappa_\star$ without solving the complicated non-linear problem
\cite{Letter}.
It is however impossible to deduce an analytical expression of 
$\kappa_\star$ from this approach and we resort 
to the following approximation. We assume that the 
micro-ion densities are slowly varying in the WS cell
so that the mean densities $n^\pm$ provide a reasonable estimation 
of the boundary densities $\rho ^{\pm}(R_{WS})$. We thus write
\be
\kappa^2_\star = 4 \pi \lb (n^++n^-) = 4 \pi \lb (2 n^+ + \Zsat	 \rho),
\ee
where the last equality follows from the electroneutrality constraint
($\rho $ is the density of colloids, assumed positively charged
without loss of generality). Note that at the level of a linear
theory, the effective charge $\Zsat$ and not the bare one $\Zbare$ 
enters this expression.

We now need to relate the mean densities $n^\pm$ to $n_0$, the concentration
in the reservoir (the so called Donnan effect,
see e.g. \cite{PHA,DesernoGrunberg}). Chemical equilibrium imposes 
$\rho^+ \rho^- = n_0^2$ at any point in the cell. We again assume this relation to hold
for the mean densities,
so that $n^+ n^- = n_0^2$. This leads to 
\be
\kappa_\star^4 = \kappa_0^4 + (4 \pi \lb \Zsat \rho)^2.
\label{eq:kap}
\ee
Up to now the reasoning is quite
general and independent of the geometry. For spheres with radius
$a$ and packing fraction $\eta = 4 \pi \rho a^3/3=(a/R_{WS})^3$,
and for rods with packing fraction $\eta=(a/R_{WS})^2$, we obtain
\be
(\kappa_\star a)^4 = (\kappa_0 a)^4 + (3 \eta \Zsat \lb/a )^2
 ~\hbox{(spheres)} \quad \hbox{and} \quad
(\kappa_\star a)^4 = (\kappa_0 a)^4 + (4 \eta \lasat\lb )^2
 ~\hbox{(cylinders)}
\label{eq:kapgeom}
\ee
Supplementing these equations with the $\kappa$ dependence of
$\Zsat$ obtained in Eq. (\ref{eq:za}) for spheres
and (\ref{eq:zb}) for cylinders, leads to an equation of degree 4 
satisfied by $\kappa_\star$ (or equivalently $\Zsat$).

This equation can be solved analytically. We give the general solution
in appendix A.
However, without added salt (formally $\kappa_0=0$), the solutions 
of these equations take a particularly simple form:
\begin{eqnarray}
&&\Zsat \frac{\lb}{a} = 6 + 24 \eta + 12 \sqrt{2\eta + 4 \eta^2}
\label{eq:zsphere}
\\
&&\lasat \lb = \frac{3}{2} + 8 \eta + 2\sqrt{6 \eta+ 16 \eta^2}
\label{eq:zcyl}
\end{eqnarray}

\begin{center}
\begin{figure}[h]
\epsfig{figure=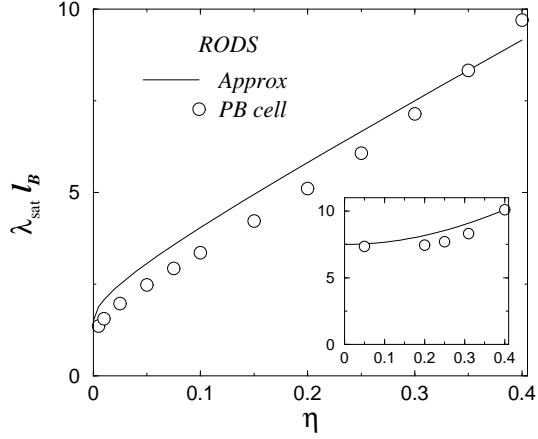,width=7cm,height=6cm,angle=0}
\caption{Effective line charge density at saturation $\lasat$ (in units of
$1/\lb$), as a function of packing fraction
for a cylindrical macro-ion of radius $a$ 
enclosed in a concentric cylindrical cell of radius $a \eta^{-1/2}$ (no added salt).
The analytical expression (\protect\ref{eq:zcyl}) is shown by the
continuous curve while the non-linear Poisson-Boltzmann values computed 
numerically following Alexander {\it et al.} \protect\cite{Alexander} 
are represented with circles. The inset shows the same quantity with added 
electrolyte ($\kappa_0 a = 3.0$).}
\label{fig:cyl}
\end{figure}
\end{center}

We compare on Figs. (\ref{fig:sph}) and (\ref{fig:cyl}),
the results obtained following this route with the full PB
estimate following Alexander's prescription. 
The latter involves a numerical resolution of the PB
equation. Both the
no-salt and the finite ionic strength case are considered.
The agreement is seen to be quite good in view of the 
minimum number of  ingredients involved in the present approach.

\section{Conclusion}
\begin{center}
\begin{figure}[h]
\epsfig{figure=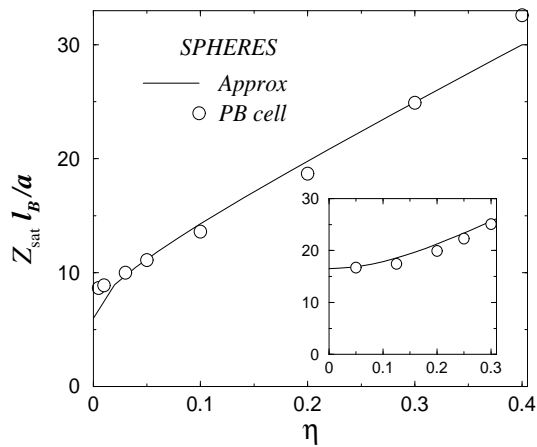,width=7cm,height=6cm,angle=0}
\caption{Effective charge at saturation $\Zsat$ (in units of
radius over Bjerrum length $\lb$), as a function of packing fraction
for a spherical macro-ion of radius $a$ 
enclosed in a concentric spherical cell of radius $a \eta^{-1/3}$.
The situation is without added salt.
The analytical expression (\protect\ref{eq:zsphere}) is shown by the
continuous curve while the non-linear Poisson-Boltzmann values computed 
numerically following Alexander {\it et al.} \protect\cite{Alexander} 
are represented with circles. Inset: same with added salt for $\kappa_0 a = 2.6$.}
\label{fig:sph}
\end{figure}
\end{center}

In this contribution, we propose a simple approximation scheme to compute 
the saturation value of the effective charge in concentrated suspensions of 
highly charged colloidal particles. 
Our estimate starts with an asymptotic expression of the effective charge 
in the infinite dilution limit, obtained from recent analytical
results \cite{Shkel}. The case of finite colloidal dilution is described using
a Donnan equilibrium approximation for the osmotic 
equilibrium of the suspension with a salt reservoir.
This calculation leads to an analytical estimate of $\Zsat$, the saturation value 
of the effective charge, as a function of the density of colloids and salt concentration.
Our starting points [Eqs. (\ref{eq:za}) and (\ref{eq:zb})] neglect
contributions of order $(\kappa_0 a)^{-1}$ and become inaccurate for 
$\kappa a <1$. This implies that our effective charges at finite density
becomes less reliable in the salt free case for small packing fractions $\eta$.
However, this is seen to occur for very small $\eta$ only (see e.g. 
Fig. \ref{fig:sph}). 

These results are obtained at the level of the mean-field Poisson-Boltzmann 
theory. We note however that the existence of a saturation value of the effective charge, $\Zsat$,
independent of the bare charge, is indeed confirmed in more refined approaches
in the {\em colloidal limit} $a\gg\lb$, with $a$ the colloid size and $\lb$ the Bjerrum length
(see e.g. Groot \cite{Groot} using the primitive model).
These results show moreover that PB theory becomes successful in the afore-mentioned
colloidal limit, $a\gg \lb$ \cite{Groot,Stevens}. Eventually the saturation picture within
PB theory is in quantitative agreement with experimental data for the osmotic pressure
\cite{Reus,BDNA,Letter}. 



It is eventually instructive to reconsider the results reported by Alexander
{\it et al.} in their original paper \cite{Alexander}. At a packing fraction $\eta = 0.125$,
they find numerically a saturation value for $\Zeff$ of the order of 15, 
in $a/\lb$ units (spherical colloid).  On the other hand our expression
(\ref{eq:zsphere}) for the same $\eta$ gives
$\Zsat \lb/a = 9 + 3 \sqrt{5}\simeq 15.7$, i.e. very close to the
value found in \cite{Alexander}. 
Note that this number of $15$ 
has subsequently been often quoted in the literature as a 
``standard'' value of the effective charge. 
The saturation value of the effective charge however
crucially depends on the colloid volume fraction, and the full
density dependence of $\Zsat$ should be taken into account in finite concentration cases.

\medskip

Acknowledgments: We would like to thank M. Deserno and 
H.H. von Gr\"unberg for useful discussions.






\begin{appendix}
\section{Analytical expression for the effective charge} 
\label{appendix} 

We give
here the explicit expression for the effective charge as a function of the colloid
packing fraction $\eta$ and salt concentration in the reservoir, $n_0= \kappa_0^2/8\pi \ell_B$.  

Inserting Eqs. (\ref{eq:za}) or (\ref{eq:zb}) into Eq.  (\ref{eq:kapgeom}), one gets
the quartic equation for $X=\kappa_\star a$~:
\be
X^4 = X_0^4 + \eta^2(\alpha + \beta X)^2
\label{eq:eq4}
\ee
with $X_0=\kappa_0 a$. Values for $\alpha$,$\beta$ are respectively
$\alpha=12$ and $\beta=18$ for spheres, and $\alpha=8$ and $\beta=6$ 
for cylinders.

Let us introduce the following quantities :
\begin{eqnarray}
&\Lambda =&-12 X_0^4-12 \beta^2 \eta^2+\alpha^4 \eta^4 \nonumber \\
&\Gamma =&108 \alpha^2 \beta^2 \eta^4-2 \alpha^6 \eta^6-72\alpha^2 \eta^2 
(X_0^4+\beta^2\eta^2) \nonumber \\
&\Phi =&-72 X_0^4 \alpha^2 \eta^2+36 \alpha^2\beta^2\eta^4-2\alpha^6\eta^6
\nonumber \\
&\Psi =& {1\over {3~2^{1/3}}} \biggl( \Phi+\sqrt{\Gamma^2-4 \Lambda^3}\biggr)^{1/3}
\nonumber \\
&\Theta=&{\Lambda\over{9 \Psi}}+\Psi
\label{defs}
\end{eqnarray}

The solution $X=\kappa_\star a$ is then found as 
\begin{eqnarray}
&\kappa_\star a=&{1\over 2} \sqrt{{2\alpha^2\eta^2\over 3}+\Theta}+{1\over 2}  \biggl(
{4 \alpha^2\eta^2\over 3}-\Theta
+{{ 4 \alpha \beta \eta^2}\over{\sqrt{{2\alpha^2\eta^2\over 3}+\Theta}}}\biggr)^{1/2}
\label{eq:sol}
\end{eqnarray}

The effective charge is then obtained by replacing this value for $\kappa_\star$ in
Eq. (\ref{eq:za}) for spherical macro ions, and in Eq. (\ref{eq:zb}) for rod-like macro ions.

In the no-added salt case, these expressions reduce to Eqs. (\ref{eq:zsphere}) and
(\ref{eq:zcyl}), as indicated in the
text.

\end{appendix}

\end{document}